\documentclass[12pt, amsmath]{revtex4-1}
\usepackage{hyperref} 
\usepackage{subfig}
\usepackage{mathrsfs} 
\usepackage{xfp} 
\usepackage{graphicx}
\usepackage{amsmath}
\usepackage{cleveref}
\usepackage{xcolor}
\usepackage{ulem} 
\usepackage{float}
\usepackage{mathtools}

\input{CQGformat.input}

\newcommand\lam{\lambda}
\newcommand\EN{\mathcal{E}}
\newcommand\ANG{\mathcal{L}}

\DeclareMathOperator{\sn}{\mathsf{sn}}

\DeclareMathOperator{\am}{\mathsf{am}}
\DeclareMathOperator{\amz}{\xi_z}
\DeclareMathOperator{\snr}{sin(\xi_r)}
\DeclareMathOperator{\cnr}{cos(\xi_r)}
\DeclareMathOperator{\amr}{\xi_r}

\newcommand{\elK}{\operatorname{\mathsf{K}}}
\newcommand{\elE}{\operatorname{\mathsf{E}}}
\newcommand{\elF}{\operatorname{\mathsf{F}}}
\newcommand{\elPi}{\operatorname{\mathsf{\Pi}}}

\begin{document}
\title{Kerr-fully Diving into the Abyss:\newline Analytic Solutions to Plunging Geodesics in Kerr}
\author{Conor Dyson$^1$,  Maarten van de Meent$^{1,2}$}

\address{$^1$ Niels Bohr International Academy, Niels Bohr Institute, Blegdamsvej 17, 2100 Copenhagen, Denmark}
\address{$^2$ Max Plank Institute for Gravitational Physics (Albert Einstein Institute), Potsdam-Golm, Germany}

\begin{abstract}
We present closed-form solutions for the generic class of plunging geodesics in the extended Kerr spacetime using Boyer-Lindquist coordinates. We also specialise to the case of test particles plunging from the innermost precessing stable circular orbit (ISSO) and unstable spherical orbits. We find these solutions in the form of elementary and Jacobi elliptic functions parameterised by Mino time. In particular, we demonstrate that solutions for the ISSO case can be determined almost entirely in terms of elementary functions, depending only on the spin parameter of the black hole and the radius of the ISSO. Furthermore, we introduce a new equation that characterises the radial inflow from the ISSO to the horizon, taking into account the inclination. For ease of application, our solutions have been implemented in a Mathematica package that is available as part of the {\tt KerrGeodesics} package in the Black Hole Perturbation Toolkit.
\end{abstract}
\section{Introduction}
Analysing the geodesic structure of a spacetime is one of the foundational means of understanding it in its unperturbed state. This is apparent in the fact that the geodesics of the Kerr spacetime have been extensively studied since its original derivation~\cite{Kerr:1963ud}. A critical step in the calculation of the geodesics in Kerr was the discovery of a fourth constant of motion, the Carter constant $Q$~\cite{Carter:1968rr}, which along with the mass shell condition, the conserved energy $\EN$ and the angular momentum $\ANG$ allow for the full integrability of the geodesic equations of motion. While exact solutions for various special cases had been derived (e.g, see~\cite{Chandrasekhar:579245} or~\cite{Wilkins:1972rs}) no real effort was made to tackle the generic case until the start of the 21st century~\cite{Kraniotis:2004cz}. Without getting explicit solutions for the geodesics themselves,~\cite{Schmidt:2002qk}~found exact expressions for the orbital frequencies with respect to coordinate time of generic bound orbits. The introduction of the Mino time parameterisation~\cite{Mino:2003yg} subsequently allowed for the full decoupling of the problem in a much simpler manner to previous approaches~\cite{Carter:1968rr}.
This opened the door for finding analytic solutions to generic bound geodesics in Kerr~\cite{Fujita:2009bp} as a system of piecewise smooth functions, which were then notably simplified through analytic continuation~\cite{vandeMeent:2019cam}.

The full class of analytic solutions to  Kerr-de Sitter and Kerr-anti-de Sitter space-times has been found generically~\cite{Hackmann:2010zz}. However these solutions are presented in the form of Weierstrass elliptic and hyperelliptic Kleinian functions, which can make them cumbersome to deal with. This work also provided part of the motivation for deriving a more explicit solution for null geodesics in the exterior of Kerr in terms of Jacobi elliptic functions~\cite{Gralla:2019ceu}. Recent work has subsequently derived these equations for the specified case of bound null geodesics~\cite{Omwoyo2022:2212.10492v1}, relevant in pursuit of forming tight constraints on Black Hole parameters by observations of the photon ring using the Event Horizon Telescope~\cite{broderick2022photon}.

Recently,~\cite{Mummery:2022ana} has derived a much simplified analytic solution for the  special case of equatorial plunging timelike geodesics which asymptote from the innermost stable circular orbit (ISCO). In this work, they also provide a simple expression for the equatorial radial inflow from the ISCO relevant to the study of accretion disk dynamics. In practice, these systems will not always be confined to the equatorial plane, motivating the generalisation of this result to the inclined (i.e. non-spin aligned or precessing) case, as we will do in this paper. In the interest of completeness we also provide solutions for plunges starting from a general (inclined and eccentric) last stable orbit (LSO), i.e. asymptoting from a generic unstable spherical orbit (USO).

Our work on generic plunges is further motivated by the fundamental role that geodesics play in the gravitational self-force approach to solving the relativistic dynamics of binary black holes~\cite{Barack:2018yvs}. In this approach the  dynamics are expanded in powers of the mass-ratio between to two black holes. In this scheme, the zeroth approximation to the motion of the lighter secondary component is given by a geodesic in the Kerr geometry generated by the (heavier) primary black hole. At higher orders, this motion is corrected by an effective force term, the gravitational self-force, causing the system to evolve.
During the inspiral phase this evolution can be solved using a 2-timescale formalism~\cite{Hinderer:2008dm,
Pound:2021qin, Miller:2020bft}, adiabatically evolving the system along a sequence of bound geodesics. The 2-timescale formalism breaks down as the system approaches the LSO around the black hole, where it enters a transition regime governed by a new intermediate timescale~\cite{Buonanno:2000ef, Ori:2000zn, OShaughnessy:2002tbu, Sperhake:2007gu, Kesden:2011ma}. In the asymptotic regime beyond the transition, the motion is again well described by a perturbed geodesic, now of the plunging variety. The timescale for the plunge is much shorter than radiation reaction timescale associated with the self-force. Consequently, the dynamics in this regime are dominated by the geodesic term. In practice, full inspiral-transition-plunge trajectories are formed by asymptotically matching an inspiral trajectory and a plunging geodesic to the jump obtained from analysing the transition regime~\cite{Ori:2000zn,Apte:2019txp}.

Development of the gravitational self-force formalism was originally motivated by the need for producing accurate gravitational waveforms for the observation of extreme mass-ratio inspirals (EMRIs) with the planned space-based gravitational wave observatory LISA. EMRIs are expected to have mass-ratios of order $10^{-5}$, and will therefore spend hundreds of thousands of orbits in the strong field regime of the inspiral phase. Consequently, the handful of orbits represented by the transition and plunge phases are generally expected to provide a negligible contribution to the total signal. However, over time, evidence has mounted~\cite{LeTiec:2011bk,Sperhake:2011ik,LeTiec:2011dp,Nagar:2013sga,LeTiec:2013uey,LeTiec:2017ebm,vandeMeent:2016hel,vandeMeent:2020xgc,Wardell:2021fyy,Ramos-Buades:2022lgf} suggesting that the self-force formalism can produce accurate results at much higher mass-ratios, and possibly even for comparable mass binaries. In these regimes the plunge and associated ringdown form a much more significant portion of the waveform. Consequently, the realisation that waveforms from the self-force formalism may be usable in the more comparable mass regime has led to a renewed interest in modelling the transition and plunge phases~\cite{Hadar:2009ip,dAmbrosi:2014llh,Compere:2021zfj, Apte:2019txp,Burke:2019yek, Compere:2019cqe, Compere:2021iwh, Compere:2021zfj} and the gravitational waves produced during the plunge and subsequent ringdown~\cite{Folacci:2018cic, Hughes:2019zmt, Lim:2019xrb}. So far this effort has focused mostly on special cases involving quasi-circular (possibly precessing) inspirals. As the work progresses towards fully generic inspirals, there is a need for generic solutions for the plunge geodesics. This paper provides the latter by solving for generic plunging geodesics in an easy to evaluate form. 

The layout of this paper is as follows, in section \ref{sec:GeodesicEquations} we begin with an introduction on the geodesic equations of Kerr and provide explicit definitions for the related conserved quantities. In section \ref{sec:ISSO} we then focus our attention on the special case of plunging timelike geodesics which asymptote from the innermost stable precessing circular orbit (ISSO).\footnote{In this paper, we use the acronym ISSO to refer to the general case of the last stable circular orbit for spherical (i.e. inclined precessing) orbits. The term ISCO will be used exclusively for the special case of circular equatorial last stable orbits.} From these equations we determine the exact and two approximate expressions for the rate of radial inflow from the ISSO to the horizon. We then go on to determine the fully analytic solutions to these geodesic equations for ISSO plunges. In section \ref{sec:deeplybound} of this work, we determine novel, fully analytic, expressions for generic timelike plunging geodesics in the Kerr spacetime, in terms of elementary and (Jacobi) elliptic functions. These generic plunges are presented in a manifestly real form such to be easily implementable, supporting current work in the self-force community on the aforementioned transition to plunge. In relation to solutions of special classes of geodesics  we also provide the solutions for plunges asymptoting from an USO in \ref{A:USO}. Finally, we have implemented these solutions in the {\tt KerrGeodesics} package of the Black Hole Perturbation Toolkit. We work in geometric units where G = c = 1.

\section{Geodesic Equations}\label{sec:GeodesicEquations}
We work in modified Boyer-Lindquist coordinates $(z = \cos(\theta))$ and denote the mass and spin of the black hole as $M =1$ and $a$, respectively. We further use the standard definitions, $\Sigma =(r^2+a^2z^2)$ and $\Delta = (r-r_{+})(r-r_{-})$. The inner and outer event horizons are given by
\begin{align}
    r_{\pm} & \coloneqq 1 \pm \sqrt{1-a^2}.
\end{align}
The symmetries of the Kerr geometry provide two constants of motion,  the conserved energy  and angular momentum, which are given as
\begin{align}
    \EN &\coloneqq -u^{\mu}g_{\mu\nu}\left(\frac{\partial}{\partial t}\right)^{\nu}, \text{ and}\\
    \ANG &\coloneqq  u^{\mu}g_{\mu\nu}\left(\frac{\partial}{\partial \phi}\right)^{\nu}, 
\end{align}
respectively. Here $g_{\mu\nu}$ is the Kerr metric tensor in modified Boyer-Lindquist coordinates. There also exists a third constant of motion $Q$ known as the Carter constant~\cite{Carter:1968rr} which arises in some families of spacetimes exhibiting Type D symmetry~\cite{Walker:1970un}. These conserved quantities arise as a result of the existence of certain Killing tensors, which are defined to satisfy the condition $\nabla_{(a}\mathcal{K}_{bc)} = 0$. In Kerr, a tensor satisfying this conditions can be found, and is given by,
\begin{align}
    \mathcal{K}_{\mu\nu} &\coloneqq  2\Sigma l_{(\mu} n_{\nu)} + r^2 g_{\mu\nu}.
\end{align}
Here $l_{\mu}$ and $n_{\nu}$ are the principal null vectors of the Kinnersly tetrad~\cite{Kinnersley:1969zza} defined by $l^{\mu} = [\frac{r^2+a^2}{\Delta}, 1 , 0, \frac{a}{\Delta} ]$, and $n^{\nu} = \frac{1}{ 2\Sigma}[r^2+a^2, -\Delta , 0, a ] $. The Carter constant is then given by, 
\begin{align}
    Q\coloneqq  u^{\mu}\mathcal{K}_{\mu\nu}u^{\nu} - (\ANG-a\EN)^2.
\end{align}
The Kerr metric also satisfies the Killing tensor condition, giving rise to a final constant of motion, $g_{\mu \nu}u^{\mu}u^{\nu} = -1$, also known as the mass shell condition,. Taking these four conserved quantities the equations of geodesic motion in Kerr are given by~\cite{Carter:1968rr},
\begin{align}
    &\begin{aligned}\label{eq:radialeom}
        \left(\frac{dr}{d\lambda}\right)^2 
        &= (\mathcal{E}(r^2+a^2)-a\mathcal{L})^2-\Delta(r^2+(a\mathcal{E}-\mathcal{L})^2+Q)\\
        &= (1-\EN^2)(r_1-r)(r_2-r)(r_3-r)(r-r_4) \\
        &= R(r), 
    \end{aligned}\\
    &\begin{aligned}\label{eq:polareom}
        \left(\frac{dz}{d\lambda}\right)^2 &= Q-z^2(a^2(1-\EN^2)(1-z^2) + \mathcal{L}^2+Q)\\
        &= (z^2-z_1^2)(a^2(1-\EN^2) z^2-z_2^2)\\
        &= Z(z),
    \end{aligned}\\
    &\;\;\begin{aligned}\label{eq:timeeom}
        \frac{dt}{d\lambda} &= \frac{(r^2+a^2)}{\Delta}(\mathcal{E}(r^2+a^2)-a\mathcal{L})-a^2\mathcal{E}(1-z^2)+a\mathcal{L},\;\;\; \text{and}
    \end{aligned}\\
    &\;\;\begin{aligned}\label{eq:azimuthaleom}
        \frac{d\phi}{d\lambda} &= \frac{a}{\Delta}(\mathcal{E}(r^2+a^2)-a\mathcal{L})+\frac{\mathcal{L}}{1-z^2}-a\mathcal{E},
    \end{aligned}
\end{align}
where we have specialised to working in Mino(-Carter)~\cite{Mino:2003yg} time defined by 
\begin{equation}
    d\tau = \Sigma d\lam.
\end{equation}
By taking advantage of the Mino time parameterisation the equations of motion completely decouple and can be solved hierarchically. This is done by first solving for $r(\lam)$ and $z(\lam)$  then naturally solving the equations in the form $t(r,z,\lam) = t_r(r)+t_{z}(z) - a \EN \lam$ and $\phi(r,z,\lam) = \phi_r(r)+\phi_{z}(z) + a\ANG\lam $. The solutions for $r(\lam)$ and $z(\lam)$ can be directly substituted to obtain $t(\lam)$ and $\phi(\lam)$. 

Analysing the radial equation in its explicit form \eqref{eq:radialeom}, one can see that the distinction between bound and plunging orbits is fully determined by the root structure of the fourth order polynomial $R(r)$. In particular, plunges occur when we have bound motion between some roots $r_i,r_j$ of $R(r)$ with $r_i<r_{-} < r_{+} < r_j$. In this work, we restrict to geodesics with $\EN<1$, which implies that the radial potential $R(r)$ is negative in the limit $r\to\infty$, ensuring that the geodesic is bound to the Kerr black hole. Moreover, since $R(r_{\pm})>0$ it guarantees that there is at least one real root of $R(r)$ outside $r_+$. A less obvious implication comes from the polar equation~\eqref{eq:polareom}. Rewriting the polar potential as
\begin{equation}
    Z(z) = (1-z^2)Q-z^2(a^2(1-\EN^2)(1-z^2) + \mathcal{L}^2),
\end{equation}
it becomes apparent that when $\EN<1$ the polar equation has solutions with $-1<z<1$ if and only if $Q\geq 0$. For the radial potential this implies that $R(0)\leq 0$, and consequently that there exists at least one real root of the radial equation between $r=0$ and the inner horizon. We thus find that for any values of $\EN<1$ and $Q\geq 0$, there exists a plunging geodesic. It is important to note that the case of $\EN>1$ with $Q<0$ also contributes to a small subset of parameter space for which a real solution is allowed. Its dynamics are quite interesting as by a brief analysis of the equations of motion one can see that this would give a test particle which plunges from infinity, poloidally oscillating around some $z_0 \neq 0$.

\begin{figure}[t!]
    \includegraphics[width=120mm]{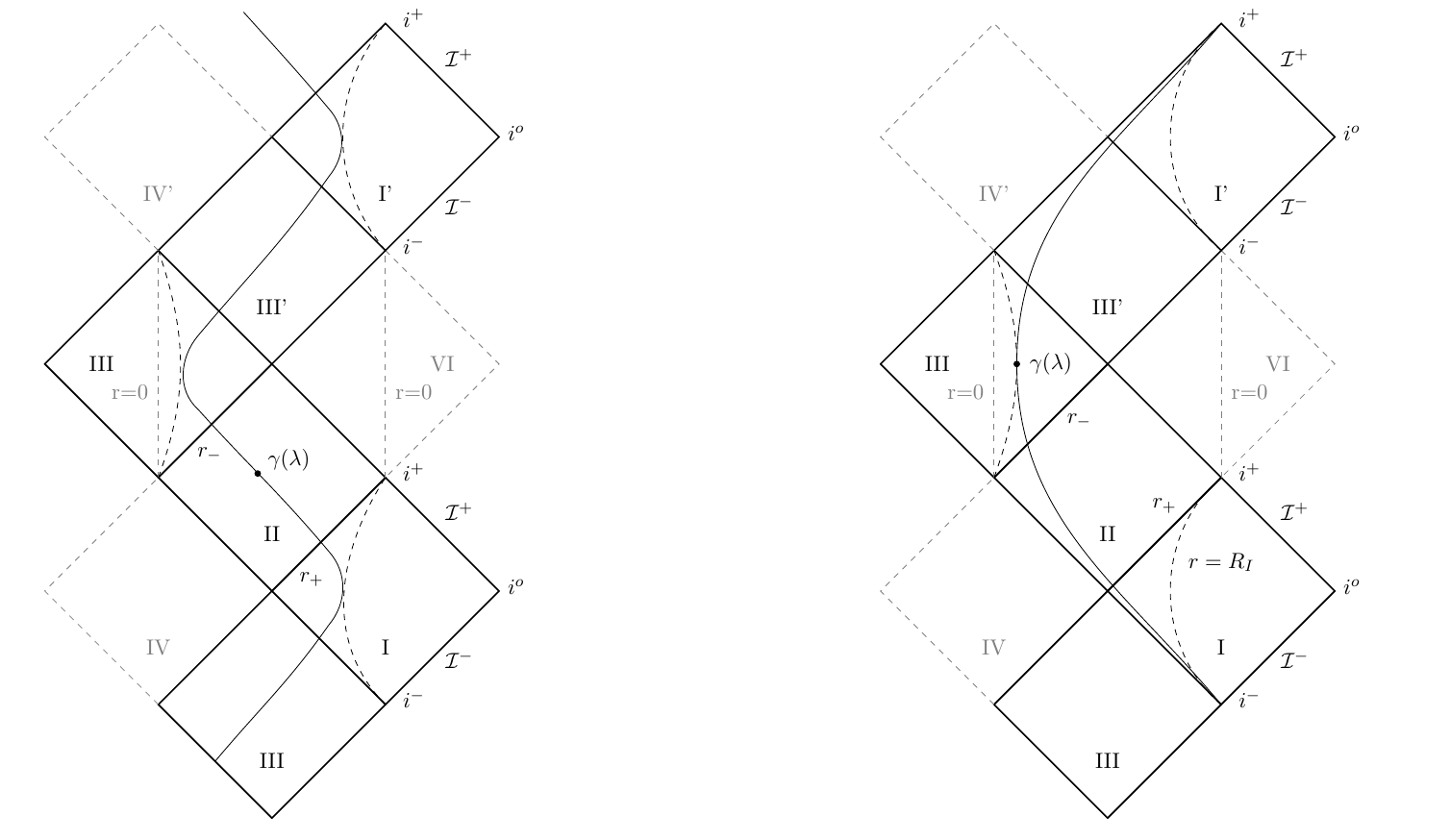}
    \caption{Penrose diagrams of the maximally analytic extension of the Kerr metric showing different plunge trajectories. Left Image: A generic plunging orbit whereby the solutions oscillate between the largest root of $R(r)$ which is less than $r_-$ and the smallest root of $R(r)$ greater than $r_+$. These trajectories correspond to geodesic motion entering the black hole, reaching a turning a point, and exiting the white hole into a different asymptotically flat region. Each point on the dashed lines in region $I$  and $I'$ correspond to a spacelike 2-surface of constant radius corresponding to the outer turning point. Right Image: A plunge which asymptotes to $r_I$ as $\lam\rightarrow \pm\infty$, where $\gamma(\lam)$ represents the geodesic in question. On the Penrose diagram these trajectories begin at past spacial infinity $i^{-}$ and end at future spacial infinity $i^{+}$ in a different asymptotically flat region. Each point on the dashed lines in region $I$  and $I'$ correspond to a spacelike 2-surface of constant radius corresponding to the ISSO radius $r_I$}
     \label{fig:penrose}
\end{figure}

Generically, a plunging geodesic will eternally oscillate between two turning points of the radial potential and return to the same radial point after a finite amount of Mino (and proper) time. Geometrically, this corresponds to a geodesic diving into the Kerr black hole and passing through the two horizons before being scattered back out, passing the horizons in reversed order and exiting in a different asymptotically flat region of the maximally extended Kerr solution, as shown in the Penrose diagrams in Fig.~\ref{fig:penrose}.

Exceptions to this picture occur for equatorial trajectories and when one of the bounding roots has a multiplicity greater than one. In the first case $Q=0$, one finds that $r=0$ is a root of the radial equation. If this is the inner turning point of the geodesic, the plunge will end on the singularity after a finite amount of Mino (and proper) time. In the second case, approaching a root with higher multiplicity takes an infinite amount of Mino (and proper) time. In Section~\ref{sec:ISSO}, we will see the physically relevant case where the outer turning point of the solution is a triple root, i.e. lies on the ISSO, where the radial solution takes a particularly simple form generalising the result of~\cite{Mummery:2022ana}. The right panel in Fig.~\ref{fig:penrose} shows this edge case trajectory in a Penrose diagram. In \ref{A:USO}, we also treat the special case where the outer turning point is a double root, generalising some of the results in~\cite{Mummery:2023hlo}.

In the generic case, we know at least two of the four roots of $R(r)$ are real. The other two roots are either both real or both complex. If the other two roots are real, they come in a pair that lies either entirely outside the outer turning point, entirely between the inner turning point and $r=0$, or entirely in the $r<0$ region. In the first of these cases, the plunging orbit exists inside of a normal bound orbit, a case sometimes referred to as  a ``deeply bound'' orbit. The solutions for cases with 4 real roots turn out to be a straightforward generalisation from the solutions of~\cite{Fujita:2009bp,vandeMeent:2019cam}. The derivation of the solution for the complex case will turn out to be substantially more involved.

\section{The Innermost Precessing Stable Circular Orbit}\label{sec:ISSO}

Before determining the solutions to fully generic plunges in Kerr we begin by solving for plunges which asymptote to the ISSO. In this case the radial potential $R(r)$ is imbued with a triple root at the ISSO radius $r=r_I$. Two of these roots come from the fact that the ISSO must be a (precessing) circular orbit and the additional root arises from the fact we are looking specifically at the innermost of these orbits meaning $R(r)$ must also inflect at this point. As a result we determine a radial equation of the form 
\begin{align}\label{eq:ISSOradialeom}
        \left(\frac{dr}{d\lambda}\right)^2 
        &=(1-\EN^2)(r_I-r)^3(r-r_4).
\end{align}
By equating Eq.~\eqref{eq:radialeom} with Eq.~\eqref{eq:ISSOradialeom} one immediately obtains the result,
\begin{equation}
    r_4 = \frac{a^2Q}{(1-\EN^2)r_{I}^{3}}.
\end{equation}
The roots of the polar equation can be found to be given by~\cite{Gralla:2019ceu},
\begin{subequations}\label{Equations of Motion}
    \begin{align}
        z_1 &= \sqrt{\frac{1}{2}\left(1+\frac{\mathcal{L}^2+Q}{a^2(1-\EN^2)}-\sqrt{\left(1 + \frac{\mathcal{L}^2+Q}{a^2(1-\EN^2)}\right)^2-\frac{4Q}{a^2(1-\EN^2)}}\right)}, \text{ and}\\
         z_2&=\sqrt{ \frac{a^2(1-\EN^2)}{2}\left(1+\frac{\mathcal{L}^2+Q}{a^2(1-\EN^2)}+\sqrt{\left(1 + \frac{\mathcal{L}^2+Q}{a^2(1-\EN^2)}\right)^2-\frac{4Q}{a^2(1-\EN^2)}}\right)}.
    \end{align}
\end{subequations}
Finally, we define
\begin{equation}
        k_z = a\sqrt{(1-\EN^2)}\frac{z_1}{z_2},
\end{equation}
as a quantity which will recurrently show up throughout this work.\footnote{Note that the definition of $k_z$ (and later $k_r$) differs from the conventions used in~\cite{vandeMeent:2019cam}.}

\subsection{Determining the Conserved Quantities}
Naturally geodesics asymptoting to the ISSO must also share the same constants of motion as the ISSO. We can therefore identify a plunging geodesic of this type with those of a particular ISSO, which has two degrees of freedom. These two degrees of freedom can be set by picking a black hole spin ($a$) and maximum orbital angle of inclination $\theta_{max} \in (-\frac{\pi}{2},\frac{\pi}{2})$. This then determines a unique $r_I$. One can also invert this relation to set the parameterisation in terms of $(a,r_I)$. Making this choice one finds for each value of $a$ there exists a range of allowed $r_I$'s each of which corresponds to a unique inclination either in prograde or retrograde. The innermost and outermost of these quantities correspond to the equatorial ISCOs in prograde and retrograde respectively. Defining $A=(27-45a^2+17a^4+a^6+8 a^3(1-a^2))^{\frac{1}{3}} $ and $B = \sqrt{3 + a^2 + \frac{9-10a^2+a^4}{A} +A} $ the range of possible $r_{I}$ values for a given $a$ are,
\begin{align}\label{eq:RIminmax}
    R_{I,\rm min/max} = 3 + &B \mp \frac{1}{2}\sqrt{\left( 72+8(a^2-6)-\frac{4(9-10a^2+a^4)}{A}-4A+\frac{64a^2}{B} \right).}
\end{align} 
Forms of these bounds have been known in the literature for some time~\cite{Bardeen:1973tla}. If one wishes to parameterise by inclination, one can use the {\tt KerrGeodesics} package in the Black hole perturbation theory toolkit~\cite{BHPToolkit} to find $\EN,\ANG$ and $Q$ parameterised by $(a,\theta_{inc})$, where $\theta_{inc}$ runs from $0$ for equatorial prograde orbits to $\pi$ for equatorial retrograde orbits~\cite{Drasco:2005kz}. In parameterising the conserved quantities by $(a,r_I)$ we begin with the expression for the marginally stable spherical orbits $Q$ written in terms of $r_I$~\cite{Teo:2020sey} which gives,
\begin{align}\label{eq:carter}
    Q &= r_I^{\frac{5}{2}}\frac{(\sqrt{(r_I-r_{+})(r_I-r_{-})}-2\sqrt{r_I})^2-4a^2}{4a^2(\sqrt{(r_I-r_{+})(r_I-r_{-})}+\sqrt{r_I}-r_I^{\frac{3}{2}})}.
\end{align}
Equating the remaining coefficients between Eq.~\eqref{eq:radialeom} and  Eq.~\eqref{eq:ISSOradialeom}  we obtain the equations,
\begin{align}
    \EN &= \frac{\sqrt{a^2Q-2r_I^3+3r_I^4}}{\sqrt{3}r_I^2},\;\;\;\text{and} \label{eq:EN}\\
    \ANG &= \pm\frac{\sqrt{3a^2Q-a^2r_I^2-Qr_I^2+3r_I^4+a^2 r_I^2\EN^2-3r_I^4\EN^2}}{r_I}.\label{eq:ANG}
\end{align}
Where the $\pm$ is determined by whether or not the $r_{I}$ picked corresponds to a prograde or retrograde orbit respectively. The correct sign is determined  by the condition, 

\begin{equation}
\text{Sign} = 
\left\{
    \begin{array}{lr}
         + , & \text{if}\;\; r_I \leq \ANG_{\rm root}\\
         - , & \text{if}\;\; r_I > \ANG_{\rm root}
    \end{array}\right.,
\end{equation}
where, defining $\kappa = \sqrt{a^2-2r+r^2}$, the value of $\ANG_{\rm root}$ is given as the root of the function 
\begin{align}
\begin{aligned}\label{eq:Lroot}
        \alpha(r) = a^6 - 3 a^2 r^{\frac{5}{2}}( 3 r^{\frac{3}{2}}+  &\kappa( 6 - 2r))+r^{\frac{9}{2}}(r^{\frac{1}{2}}(20 -11r) + \kappa(5 + 3r))\\
        &+ a^4( r(3r -4) + \kappa r^{\frac{1}{2}}(1 + 3r) ),  
\end{aligned}
\end{align}
which is real and closest to $r = 6$. Eq.~\eqref{eq:Lroot} has been given in a form such that the root can be found numerically to high precision which is not the case for Eq.~\eqref{eq:ANG}. It is worth noting that as $\EN, \ANG$, and $Q$ have all been determined in terms of $r_I$, and $z_1$ is the root that defines the maximum range of oscillation allowed for a given $r_I$, our solution immediately defines a spacelike surface $(r,z,\phi) = (r_I,z_1(r_I), u)$ for $r_I \in (r_{Imin}, \ANG_{\rm root})$ and $u\in(0,2\pi)$ inside which no stable spherical orbits can exist. Taking our exact solution for this spacelike surface to the extremal limit also provides us with the ISSO surfaces found in the near-horizon extremal Kerr geometry previously found in both~\cite{Compere:2020eat} and~\cite{Stein:2019buj}.
 
\subsection{Radial Inflow}
We are now ready to begin analysing the physical consequences of extending the results of~\cite{Mummery:2022ana}, regarding equatorial ISCO flow to inclined orbits. The exact solution to this equation includes functional dependence on certain Jacobi elliptic functions. In order to simplify the results we provide two approximate forms of the inflow. The first, approximated to a modified form of the equatorial inflow equation which removes all dependence on Jacobi elliptic functions. The second, a polar averaged form which simplifies the functional dependence on the Jacobi elliptic functions to a constant dependence for any given set of parameter values. We find the modified equatorial flow  to be given by,
\begin{equation}\label{eq:radialinflowEquatorial}
     \frac{dr}{ d t}\bigg|_{\rm Equatorial} =\frac{-\sqrt{(1-\EN^2)(r_I-r)^3(r-\frac{a^2Q}{(1-\EN^2)r_{I}^{3}})}}{a\mathcal{L}+\frac{(r^2+a^2)}{\Delta}(\EN(r^2+a^2)-a\mathcal{L})-a^2\EN}.
\end{equation}
In the equatorial limit ($Q=0$), this reproduces the result of~\cite{Mummery:2022ana}, but gives an improved representation of the radial inflow for inclined disks where $Q$, $\EN$ ,and $\ANG$ are given by their true values Eqs.~(\ref{eq:carter}),(\ref{eq:EN}) and (\ref{eq:ANG}) respectively. Next we go on to determine the form of the radial inflow when averaged over the polar period, this is found to be
\begin{equation}\label{eq:radialinflowPolarAVG}
     \frac{dr}{ d\langle t \rangle _{z}}\bigg|_{\rm PolarAvg} =\frac{-\sqrt{(1-\EN^2)(r_I-r)^3(r-\frac{a^2Q}{(1-\EN^2)r_{I}^{3}})}}{a\mathcal{L}+\frac{(r^2+a^2)}{\Delta}(\mathcal{E}(r^2+a^2)-a\mathcal{L})-  a^2 \EN+ \frac{z_2^2 \EN}{1-\EN^2}\left( 1 - \frac{\elE( k_z^2)}{ \elK(k_z^2)} \right)}.
\end{equation}
Where $\elK(\cdot)$  and $\elE(\cdot)$ are the complete elliptic functions of the first and second kind respectively. We have defined the polar average as follows, given some function $f(\lambda)$ that (partially) depends on $\lam$ through $z$ such that $f(\lambda)=F(r(\lam),z(\lam),\lam)$, we take 
\begin{equation}
\langle f \rangle _{z}(\lam) = \frac{1}{\Lambda_z}\int_{-\Lambda_z/2}^{\Lambda_z/2}F(r(\lam),z(\lam+\delta),\lam)d\delta,
\end{equation}
where
\begin{figure}[t!]
    \centering
    \includegraphics[width=120mm]{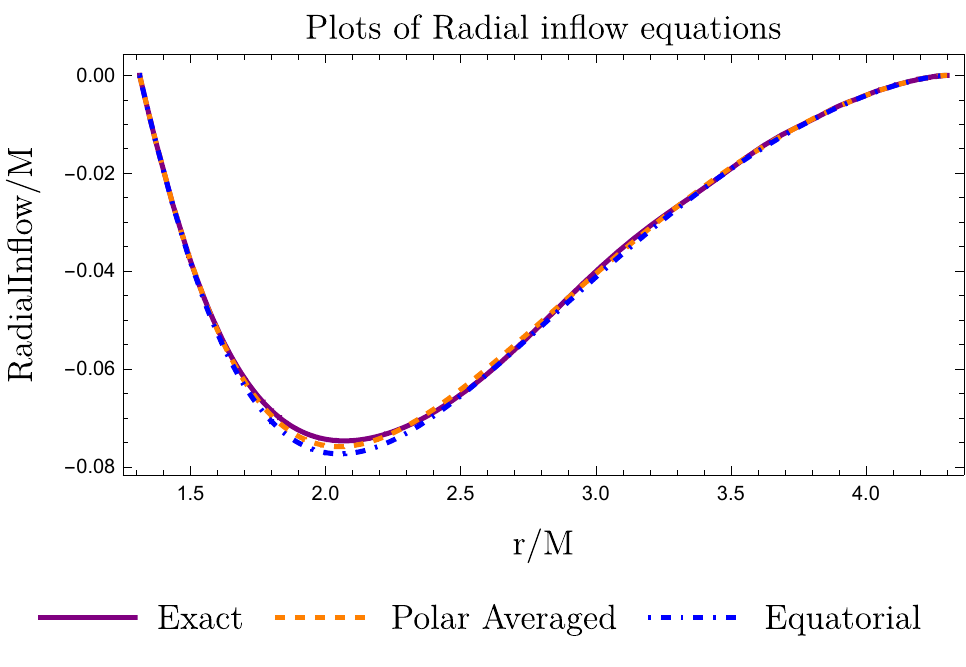}
        \caption{Plot of differing radial inflow equations from the ISSO to horizon, for parameter values $(a,r_I) = $(0.9,4) for plot range $(r_+,r_I)$ on the radial axis. The lightly oscillating purple line corresponding to the exact solution exhibiting small oscillations over each polar period. The blue and orange dashed lines correspond to the equatorial and polar averaged approximations respectively. The Radial flow measured on the y axis is given by the dimensionless quantity $dr/dt$ for each of the three flow equations.}
    \label{fig:flows}
\end{figure}
\begin{equation}
    \Lambda_z = \frac{4 \elK(k_z^2)}{z_2},
\end{equation}
is the polar period.
For example when we have an equation of the form $t(r,z,\lam) = t_r(r)+t_z(z)+a \ANG \lam$ then $\langle t \rangle _{z}(\lam)  = t_r(r(\lam))+\langle t_z \rangle _{z}(\lam) +a \ANG \lam$. Here the purpose of taking this average is to integrate out the oscillatory dependence and isolate the secular dependence in $\lam$ of terms originally containing polar dependence.
Finally, we give the exact equation for the radial inflow as, 
\begin{equation}\label{eq:radialinflowTrue}
   \frac{dr}{ dt } =\frac{-\sqrt{(1-\EN^2)(r_I-r)^3(r-\frac{a^2Q}{(1-\EN^2)r_{I}^{3}})}}{a\mathcal{L}+\frac{(r^2+a^2)}{\Delta}(\mathcal{E}(r^2+a^2)-a\mathcal{L})-a^2\EN(1 -  z_1^2 \mathrm{\sn}^2\big( \frac{2 z_2 \sqrt{r-r_4}}{\sqrt{(1-\EN^2)(r_I-r)(r_I-r_4)^2}}\big| k_z^2\big) )}.
\end{equation}
Where $\sn(\cdot|\cdot)$ is the Jacobi sine function.

\begin{figure}[t!]
    \centering
    \includegraphics[width=150mm]{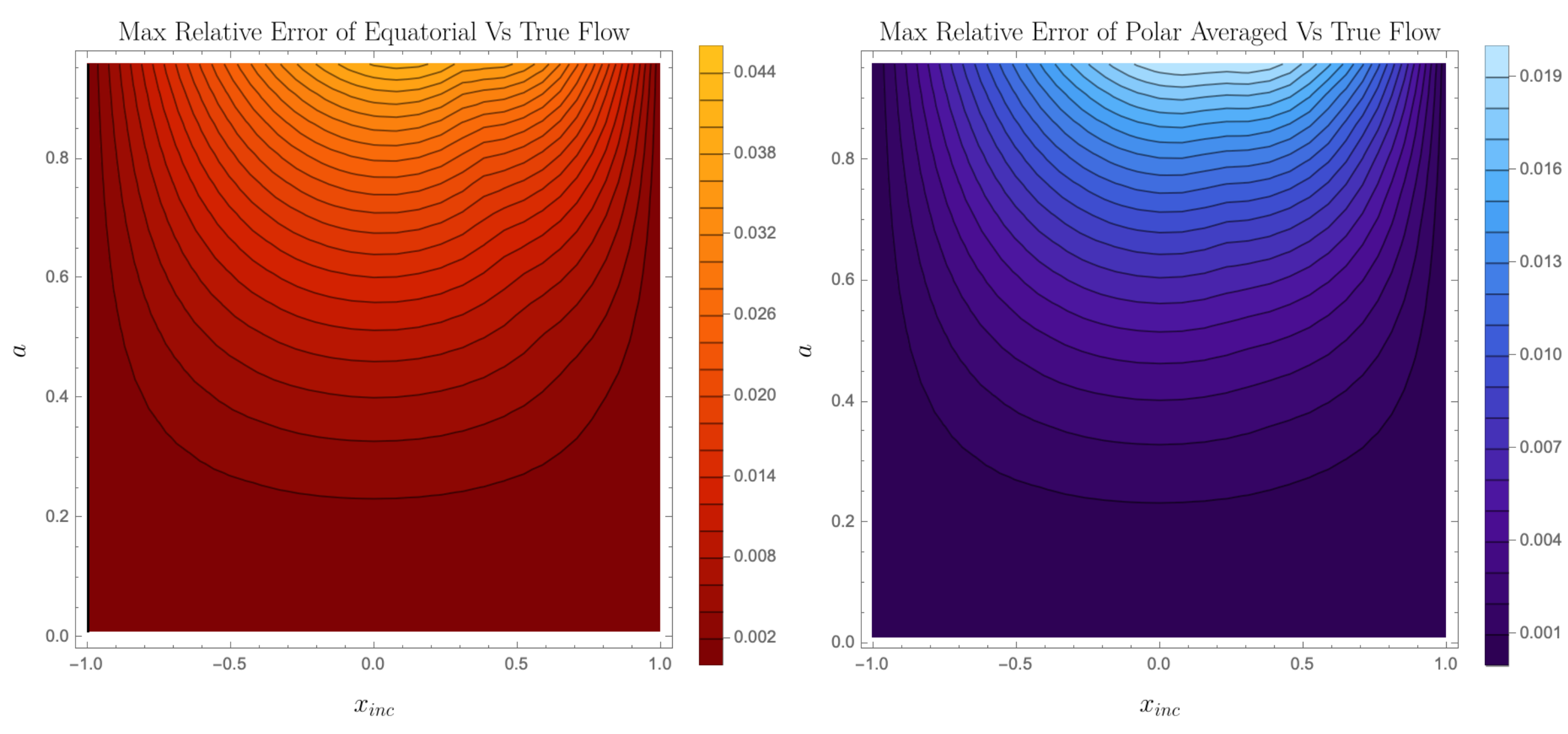}
        \caption{Left Image: Plot of the maximum relative error between the equatorial and true flow over the range $(r_+,r_I)$. Right Image: Plot of the maximum relative error between the polar averaged and true flow over the range $(r_+,r_I)$. Both plots give the error over the entire parameter space $(a,x_{inc})$.}
    \label{fig:flowerrors}
\end{figure}

In Fig.~\ref{fig:flows}, which depicts the radial flow from the differing equations, we see an improvement of accuracy in the polar averaged approximation over the equatorial approximation. We also provide a brief error analysis comparing our approximated solutions to the true solution Eq.~\eqref{eq:radialinflowTrue} over the entire parameter space. Here we parameterise our space using $(x_{inc},a)$ where $x_{inc} = \cos{\theta_{inc}}$ such that   $(x_{inc},a) \in(\{-1,1\}, \{0,1\})$. From Fig.~\ref{fig:flowerrors} we see that over the parameter space the equatorial approximation confines the error to be below $5\%$  whilst the polar averaged form provides a bound of 2$\%$ maximum relative error. In reality this is a generous upper bound and for the majority of parameter space the error of the approximated particle inflow will be notably smaller, as can be seen in Fig.~\ref{fig:flowerrors}.

\subsection{Solutions to the Equations of Motion}
At this stage we are now ready to solve the equations of motion. In the generic case we expect the solutions to arise in terms of elliptic functions. In the ISSO case however, we find that the triple root significantly simplifies the terms with radial dependence to the form of elementary functions. For convenience the solutions for the entirety of the ISSO case are given with initial conditions $(t(\lam),r(\lam),z(\lam),\phi(\lam))|_{\lam=0} =(0,r_4,0,0)$. The full solution can be easily reconstructed using Eq.~\eqref{eq:MINOISSO}.
\subsubsection{Radial Equation}
Solving Eq.~\eqref{eq:ISSOradialeom} and inverting the solution gives 
    \begin{align}\label{eq:MINOISSO}
        \lambda(r) &= \frac{2\sqrt{r-r_4}}{\sqrt{(1-\EN^2)(r_I-r)(r_I-r_4)^2}},\text{ and}\\
        r(\lambda) &= \frac{r_I(r_I-r_4)^2(1-\EN^2)\lambda^2+4r_4}{(r_I-r_4)^2(1-\EN^2)\lambda^2+4},
\end{align}
respectively.
\subsubsection{Polar Equation}
The polar solution is unchanged from the generic bound case~\cite{vandeMeent:2019cam} and given by 
\begin{equation}\label{Eq:ISSOPolar}
    z(\lambda) = z_1 \sin{\xi_z(\lam)}.
\end{equation}
Here we have defined,
\begin{equation}
    \xi_z(\lam)= \am(z_2 \lam| k_z^2),
\end{equation}
where $\am(\cdot|\cdot)$ is the Jacobi amplitude function. 
Conveniently for the remaining equations of motion, the polar and radial dependence in the time and azimuthal equation fully decouple from each other in Mino time. This means we are only required to resolve the radial component as the polar part will simply remain the same as has already been found for bound orbits in~\cite{Fujita:2009bp,vandeMeent:2019cam}. The polar dependence still remains in the form of elliptic integrals where $\elF(\cdot)$, $\elE(\cdot)$ and $\elPi(\cdot)$ are the elliptic integral of the first, second and third kind respectively.
\subsubsection{Azimuthal Equation}

Going on to solve the azimuthal component we first rewrite Eq.~\eqref{eq:azimuthaleom} as
\begin{align}
\begin{aligned}
        d\phi &= \frac{-a}{\Delta}\frac{\mathcal{E}(r^2+a^2)-a\mathcal{L}}{\sqrt{(1-\EN^2)(r_I-r)^3(r-r_4)}}dr\\
        &\qquad+\frac{\mathcal{L}}{1-z^2}\frac{1}{\sqrt{(z^2-z_1^2)(a^2(1-\EN^2) z^2-z_2^2)}}dz
        -a\mathcal{E}d\lam. 
\end{aligned}
\end{align}
We then integrate each of the terms individually. The component comprising of r dependence is found to be given by,
\begin{align}
\begin{aligned}
    \!\phi_r(\lam)& = a \frac{(\EN(r_I^2+ a^2) - a \ANG)\lam }{(r_I-r_{-})(r_I-r_{+})} + \frac{a}{\sqrt{(1-\EN^2)}}\Bigg(\frac{(\EN(r_{-}^2+ a^2) - a\ANG)}{2\sqrt{r_{-}-r_4}(r_I-r_{-})^{\frac{3}{2}}(r_{+}-r_{-})}\times \\
     & \log\left({\frac{\big(2\sqrt{r_{-}-r_4}+\lam(r_I-r_4)\sqrt{(1-\EN^2)(r_I-r_-)}\big)^2}{\big(2\sqrt{r_{-}-r_4}-\lam(r_I-r_4)\sqrt{(1-\EN^2)(r_I-r_-)}\big)^2}}\right)+ (r_{-} \Longleftrightarrow r_{+}) \Bigg),
\end{aligned}
\end{align}
where the arrow notation denotes taking the other term within the shared brackets and swapping all occurrences of $r_-$ with $r_+$. The component with $z$ dependence is then given by,
\begin{align}\label{eq:polarphi}
    \phi_z(\lam) = \frac{\ANG}{z_2}  \elPi(z_1^2; \xi_z(\lam) | k_z^2 ).
\end{align}
The polar average form of this solution can also be found and dramatically simplifies the functional dependence,
\begin{align}
    \langle \phi_z \rangle_z(\lam) = \frac{\ANG}{ \elK(k_z^2)}\elPi(z_1^2; k_z^2 ) \lam,
\end{align}
where $\elPi(\cdot,\cdot)$ is the complete elliptic integral of the third kind. Thus, we find the full azimuthal solution to be given by 
\begin{align}
    \phi(\lam) = \phi_r(\lam) + \phi_z(\lam) - a \EN \lam.
\end{align}
\subsection{Time Equation}
We perform a similar separation in the integration of the equation of motion for coordinate time giving,

\begin{align}
\begin{aligned}
    t_r(\lam) &= \frac{(a^2+r_I^2)(\EN(r_I^2+ a^2) - a \ANG)\lam }{(r_I-r_{-})(r_I-r_{+})}+ \frac{2(r_I - r_4)^2\EN \lam}{4 + (1-\EN^2)(r_I-r_4)^2\lam^2}\\
    &- \frac{(r_4+3r_I + 2(r_{+}+r_{-}))}{\sqrt{(1-\EN^2)}}\EN\arctan \left( \frac{\lam (r_I-r_4)\sqrt{(1-\EN^2)}}{2}\right)\\
    &+ \Biggl( \frac{(a^2+r_{-}^2)(\EN(r_{-}^2+ a^2) - a \ANG)}{2\sqrt{r_{-}-r_4}(r_I-r_{-})^{\frac{3}{2}}(r_{+}-r_{-})\sqrt{(1-\EN^2)}}\times \\
    & \log\bigg({\frac{\big(2\sqrt{r_{-}-r_4}+\lam(r_I-r_4)\sqrt{(1-\EN^2)(r_I-r_-)}\big)^2}{\big(2\sqrt{r_{-}-r_4}-\lam(r_I-r_4)\sqrt{(1-\EN^2)(r_I-r_-)}\big)^2}}\bigg)+ (r_{-} \Longleftrightarrow r_{+}) \Biggl).
\end{aligned}
\end{align}
Next we find the polar dependent part of the time solution to be given by,
\begin{align}\label{eq:polartime}
    t_z(\lam)  = \frac{ \EN}{1-\EN^2}\left( (z_2^2 - a^2(1-\EN^2))\lam - z_2\elE(\xi_z(\lam)|k_z^2) \right),
\end{align}
and the polar averaged form of this solution is given by, 
\begin{align}
    \begin{aligned}
   \langle t_z\rangle_z (\lam)  = \frac{z_2^2 \EN}{1-\EN^2}\left( 1 - \frac{a^2(1-\EN^2)}{z_2^2} - \frac{\elE( k_z^2)}{ \elK(k_z^2)} \right)\lam.
    \end{aligned}
\end{align}
This is the expression that was necessary to derive the polar averaged radial flow in Eq.~\eqref{eq:radialinflowPolarAVG}.
Putting this all together we then find the full time solution to be given by 
\begin{align}
    t(\lam) = t_r(\lam) +   t_z(\lam) + a \ANG \lam.
\end{align}
We find that all of the novel radial integrals which are required to be solved can be done so without too much issue through the use of partial fractions. They can then be fully analytically continued by applying some simple trigonometric substitutions. 

Additionally one can follow this approach to find the solution for proper time as a function of Mino time along a plunging ISSO geodesic,

\begin{align}
&\begin{aligned}[t]
    \tau_r(\lam) =&\left(r_I^2+\frac{2(r_I-r_4)^2}{4 + (1-\EN^2)(r_I-r_4)^2\lam^2}\right)\lam\\
    &\quad+ \frac{ (r_4^2 +2r_4r_I-3r_I^2)\arctan(\frac{\lam(r_I-r_4)\sqrt{(1-\EN^2)}}{2})}{\sqrt{(1-\EN^2)}(r_I-r_4)},
\end{aligned}
\\
    &\tau_z(\lam) =\frac{ z_2 }{(1-\EN^2)}\left( F(\amz|k_z^2 )-E(\amz|k_z^2)\right), \text{ with}
\\
    &\tau(\lam) = \tau_r(\lam)+ \tau_z(\lam).
\end{align}

\begin{figure}[tb!]
\centering
    \includegraphics[width=100mm]{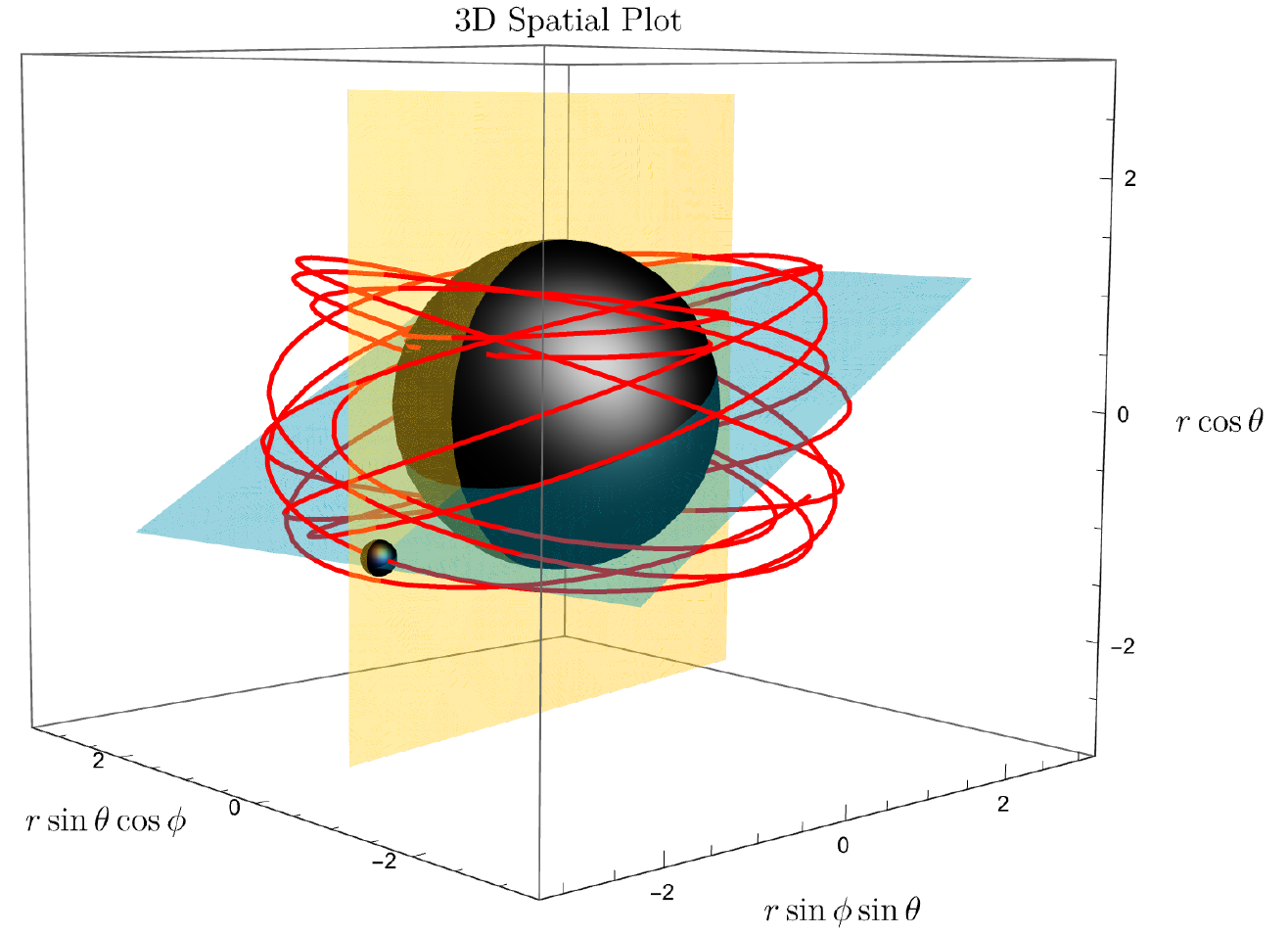}
    \caption{Orbital plots of plunging geodesics which asymptote to the ISSO in the infinite past with $(a,r_I) = (0.9,2.6)$ and $\theta = \mathrm{arccos}(z)$. Here the cyan and orange planes are the azimuthally co-precessing and poloidally co-rotating planes respectively. Examples of the projection on these planes are shown in Fig.~\ref{fig:coorotatingISSOplots}.} 
    \label{fig:ISSOplots}
\end{figure}
From Fig.~\ref{fig:ISSOplots} it can now be seen explicitly that enforcing the triple root places us in a regime where these geodesics asymptote from the ISSO and subsequently plunge in through the horizon. This showcases a number of properties of the plunge in Boyer-Lindquist co-ordinates. The structure of these geodesics is more easily depicted when orthogonally projected onto azimuthally co-precessing and poloidally co-rotating planes fixed to a particle as it follows the geodesic.
\begin{figure}[tb!]
    \centering
    \includegraphics[width=140mm]{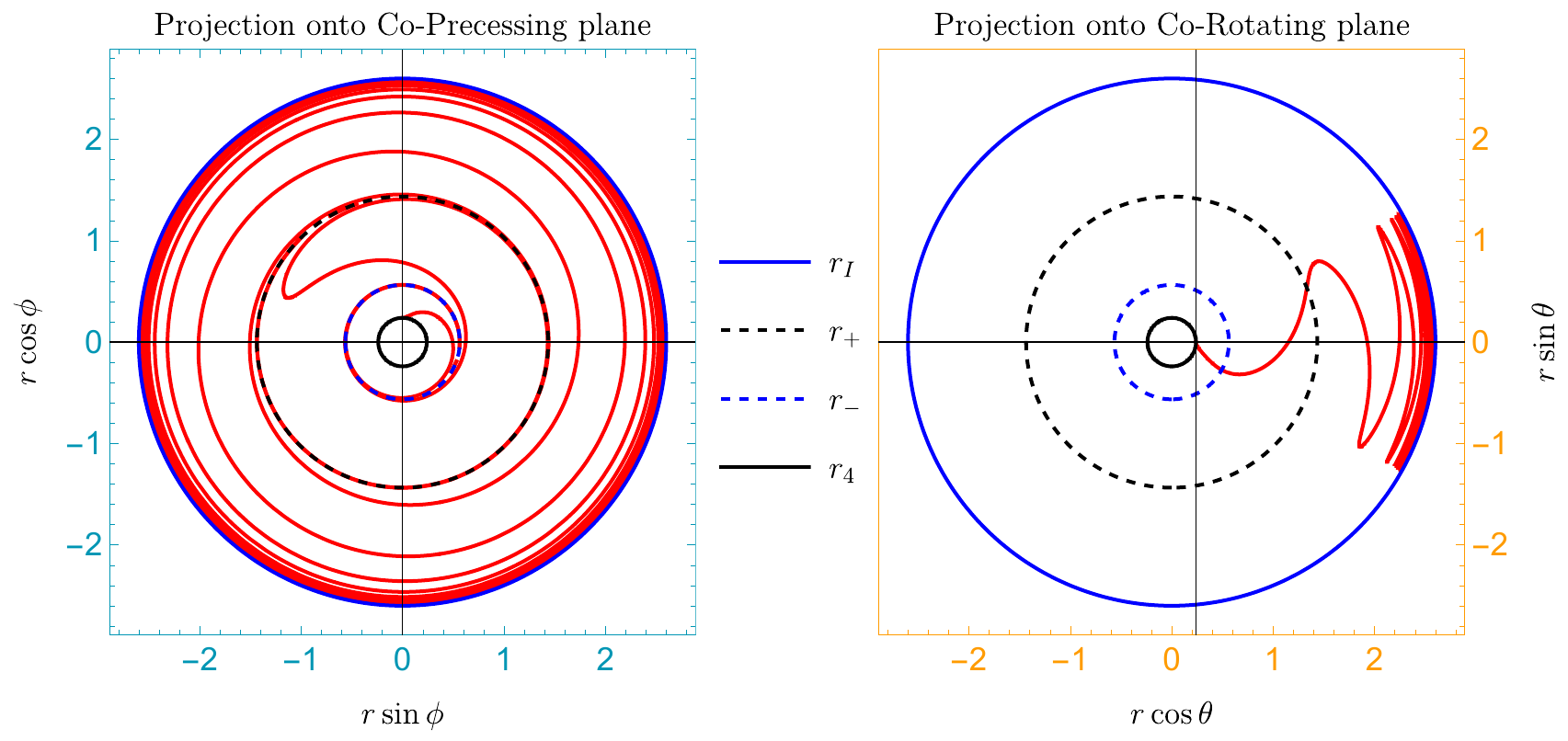}
    \caption{Co-rotating orbital plots of plunging geodesics which asymptote to the ISSO in the infinite past with $(a,r_I) = (0.9,2.6)$ . Left Image: orthogonal projection of ISSO plunge onto the co-precessing azimuthal plane. Right Image: orthogonal projection of ISSO plunge onto co-rotating polar plane. The cyan and orange frames represent the projection onto the planes seen in \ref{fig:ISSOplots} as they track a particle following the geodesic trajectory.}
    \label{fig:coorotatingISSOplots}
\end{figure}
In Fig.~\ref{fig:coorotatingISSOplots} we see that the azimuthal coordinate diverges at both the inner and outer horizons. Analysis of the coordinate time solution shows it also diverges at these points, this occurs for a similar reason as to the coordinate time divergence at the horizon in Schwarzschild coordinates for a non-spinning black hole, i.e. due to the infinite redshift. In much the same way as we have an infinite redshift on the horizon one can intuitively think of this azimuthal divergence occurring as a consequence of the infinite redshift on the horizon forcing the geodesics to also co-rotate with the black hole an infinite number of times before passing through the horizon. Naturally this is only a co-ordinate singularity and the process occurs in finite Mino and proper time. Due to this discontinuity care must be taken in the choice of branch between $r_+$ and $r_-$. The solutions depicted in  \ref{fig:ISSOplots} and \ref{fig:coorotatingISSOplots}  seem to invert between the horizons but it can be checked that our solutions conserve $\EN,\ANG$ and $Q$ throughout their parameterisation, confirming we have chosen the correct branch. In the equatorial ($Q=0$) limit, these solutions are found to agree with the results of~\cite{Mummery:2022ana}.

\section{Fully Generic Plunging Orbits}\label{sec:deeplybound}

We now move on to the case of generic timelike plunging geodesics in Kerr with no restriction on the root structure of the effective radial potential. In this regime the integrals we need to compute are notably more involved, separating into two primary classes. The first being when the radial potential admits four real roots and the second being when the radial potential admits two real and two complex roots. In the case with $R(r)$ having four real roots ($r_4<r_3<r_2<r_1$) with  $r_4<r_-<r_+<r_3<r_2<r_1$  the solution can be directly borrowed from the bound orbit case~\cite{vandeMeent:2019cam,Fujita:2009bp}  with the substitution $r_1 \Longleftrightarrow r_3$ and  $r_2 \Longleftrightarrow r_4$. In addition, in the case of 4 real roots  ($r_4<r_3<r_2<r_1$) with $r_4<r_3<r_2<r_-<r_+<r_1$ then the solutions can again be found from~\cite{vandeMeent:2019cam,Fujita:2009bp}, this time with no modification. Naively, in the case of two real and two complex roots (the interesting case for self-force plunges) one may think it to be sufficient to simply take the previously found bound solutions and analytically continue them to allow for complex values of the radial roots. With care this can be done to give correct answers, however intermediate terms in the evaluations give rise to large cancellations between complex numbers effecting numerical accuracy and evaluation speed. To find these solutions in a manifestly real, easy to evaluate, and practical manner, we are required to begin the procedure of solving the complicated elliptic integrals from scratch.

\subsection{Generic Plunge Orbits with Two Complex Roots}
In calculating the integrals for this case we follow the procedure outlined in~\cite{LabanMutrie} and give an overview of the steps involved. We begin by again noting  that only the radial components of each of the equations need to be solved as the other remaining terms are identical to those already found in the ISSO case above. Recall we now have the general expression $R(r) = (\mathcal{E}(r^2+a^2)-a\mathcal{L})^2-\Delta(r^2+(a\mathcal{E}-\mathcal{L})^2+Q)$ where $\EN$, $\ANG$ and $Q$ are all independent quantities. At this stage the integrals of concern are given by 
\begin{align}\label{Kerr EOM}
    \begin{aligned}
        \lambda &= -\int\frac{dr'}{\sqrt{ R(r')}},\\
        t_r(r) &= -\int \frac{(r'^2+a^2)(\EN(r'^2+a^2)-a\mathcal{L}) dr'}{\Delta\sqrt{ R(r')}}, \;\; \text{and} \\
        \phi_r(r) &= -\int \frac{a(\mathcal{E}(r'^2+a^2)-a\mathcal{L})dr'}{\Delta\sqrt{ R(r')}}. 
    \end{aligned}
\end{align}
Although these integrals look to be of the same form as those we just solved for in the ISSO case without much mention, the cause for the substantial increase in complexity is due to the fact that $R(r)$, in general, no longer contains any double or triple roots. This forces us to much more carefully consider the elliptic integrals at hand. The key idea in the procedure we wish to apply is to reduce the integrals of Eq.~\ref{Kerr EOM}  such that we must only solve integrals of the form,
\begin{align}\label{eq:ellipticBasis}
    \begin{aligned}
     &\lam = -\int\frac{dr}{\sqrt{ R(r)}},\;\;\; \mathcal{I}_r = \int\frac{rdr}{\sqrt{ R(r)}},\\
     &\mathcal{I}_{r^2} = \int\frac{r^2dr}{\sqrt{ R(r)}}, \;\; \text{and} \;\;\mathcal{I}_{r_{\pm}} = \int\frac{dr}{(r-r_{\pm})\sqrt{ R(r)}}.
    \end{aligned}
\end{align}
We solve these integrals in terms of the radial co-ordinate $r$, which can then be parameterised in terms of Mino time by inverting the solution for $\lam(r)$. We begin solving these integrals by first continually applying partial fractions to the radial parts of the $\phi$ and $t$ equations until arriving at the forms,
 \begin{equation}\label{eq:reducedformphi }
        \phi_r = a\left( \frac{(\EN(r_-^2+ a^2) - a \ANG)}{(r_{-}-r_{+})} \mathcal{I}_{r_{-}} + (r_{-} \Longleftrightarrow r_{+})\right) + a \EN \lam, \text{ and}
\end{equation}

\begin{align}\label{eq:reducedformt}
\begin{aligned}
    t_{r} = \EN(r_{+}^2&+r_{-}^2 + r_{+}r_{-}+2a^2)\lam + \EN\left(\mathcal{I}_{r^2} +  \mathcal{I}_{r}(r_{-} +r_{+})\right)\\
    &+\left(\frac{(r_{-}^2+a^2)(\EN(r_-^2+ a^2) - a \ANG)}{r_{-}-r_{+}}\mathcal{I}_{r_{-}} + (r_{-} \Longleftrightarrow r_{+})\right)-a \ANG \lam.
\end{aligned}
\end{align}
At this point we can now concentrate on calculating the four elliptic integrals defined in Eq.\eqref{eq:ellipticBasis}. We do this my applying a transformation given for the case of two complex roots in~\cite{LabanMutrie}. This procedure begins by letting $R(r)$ have two real roots $r_2 < r_1$ and two complex roots $r_3$ and $r_4$. We then rewrite $R(r)$ in the form $R(r) = (1-\EN^2)(r_1-r)(r-r_2)(r^2 - 2\rho_rr + \rho_r^2 -\rho_i^2) $ where  $\rho_r=\Re(r_3)$ and $\rho_i=\Im(r_4)$. Further we define
\begin{align}
    \begin{aligned}
A &= \sqrt{(r_1-\rho_r)^2+\rho_i^2} , \;\;\;B = \sqrt{(r_2-\rho_r)^2+\rho_i^2} ,\\ 
f &= \frac{4 A B}{(A-B)^2},\;\;\;k_r = \sqrt{\frac{(r_1-r_2)^2 - (A-B)^2}{4 A B }},  \text{ and }\;\;\; \\
p_2 &= r_2A^2+r_1B^2-(r_1+r_2)AB.
    \end{aligned}
\end{align}
 Next, motivated by the tables provided in~\cite{LabanMutrie}, we make the substitution in the integrals Eq.~\eqref{eq:ellipticBasis} of the form, 
\begin{equation}\label{eq:transformation}
    r(y) = \frac{p_2y^2+2(r_1+r_2)AB+2(r_1-r_2)AB\sqrt{1-y^2}}{(A-B)^2y^2+4AB}.
\end{equation}
Applying this transformation to Eq.~\eqref{eq:ellipticBasis} and again repeatedly applying partial fractions we find each integral reduces to a sum over elliptic integrands and rational polynomials. The solutions we then obtain are only analytic on the range $r\in (r_2, \frac{r_1A+r_2B}{A+B})$ where for convenience we have set the initial conditions to be given by $(t(\lam),r(\lam),z(\lam),\phi(\lam))|_{\lam=0} =(0,r_2,0,0, )$. Next we analytically extend these solutions through the point $r =\frac{r_1A+r_2B}{A+B}$ by use of trigonometric and elliptic substitutions providing a fully analytic solution on the range $r\in(r_2,r_1)$. The fully analytical solution to these integrals is then given by
\begin{equation}\label{eq:MINOsolution}
     \int\frac{dr}{\sqrt{ R(r)}} = \frac{1}{\sqrt{(1-\EN^2)A B}} \elF\left( \frac{\pi}{2}-
\arcsin\left(\frac{B(r_1-r) - A(r-r_2)}{ B(r_1-r)+A(r-r_2)}\right)\bigg| k_r^2\right),\;\;\;\;\;\;\;\;\;
\end{equation}

\begin{equation}\label{eq:Ir}
    \begin{aligned}
 \;\;\;\; \mathcal{I}_r(\lam) =\frac{Ar_2-Br_1}{A-B}\lam -\frac{1}{\sqrt{(1-\EN^2)}}&\arctan  \left( \frac{(r_1-r_2)}{2\sqrt{AB}}\frac{\snr}{\sqrt{1-k_r^2\snr^2}}\right)\\
    & + \frac{(A+B)(r_1-r_2)}{2(A-B)\sqrt{(1-\EN^2) A B}}\elPi\left( -\frac{1}{f}; \amr|k_r^2\right),
    \end{aligned}
\end{equation}
\begin{equation}\label{eq:Ir2}
    \begin{aligned}
   \mathcal{I}_{r^2}&(\lam) =\frac{(Ar_2^2-Br_1^2)}{(A-B)}\lam + \frac{\sqrt{AB}}{\sqrt{1-\EN^2}}\elE\left(\amr|k_r^2\right)\\
  &- \frac{(A+B)(A^2+2r_2^2-B^2-2r_1^2)}{4(A-B)\sqrt{(1-\EN^2) A B}}\elPi(-\frac{1}{f};\amr|k_r^2)\\
  &-\frac{\sqrt{A B}(A+B-(A-B)\cnr)}{(A-B)\sqrt{(1-\EN^2)}}\frac{\snr\sqrt{1-k_r^2\snr^2}}{(f+\snr^2)}\\
  &+\frac{A^2+2r_2^2-B^2-2r_1^2}{4(r_1-r_2)\sqrt{(1-\EN^2)}}\times\\
  \;\;\;\;\;\;\;&\arctan\left(f - (1+2 f k_r^2)\sin^2(\xi_r), 2 \sin(\xi_r) \sqrt{1-k_r^2\sin^2(\xi_r)}\sqrt{f(1+f k_r^2)}\right),
    \end{aligned}
\end{equation}  
and, 
\begin{equation}\label{eq:Irpm}
    \begin{aligned}
   \mathcal{I}_{r_{\pm}}(\lam) &= \frac{(A-B)\lam}{A(r_2-r_{\pm})-B(r_1-r_{\pm})}\;\;\;\;\;\;\;\;\;\;\\\
   &+ \frac{(r_1-r_2)(A(r_2-r_{\pm})+B(r_1-r_{\pm}))\elPi\left(\frac{1}{D_{\pm}^2}; \amr|k_r^2\right)}{2\sqrt{(1-\EN^2) A B}(r_{\pm}-r_2)(r_1-r_{\pm})(A(r_2-r_{\pm})-B(r_1-r_{\pm}))}\\
   -&\frac{\sqrt{r_1-r_2}
    }{4\sqrt{(1-\EN^2)(r_1-r_{\pm})(r_{\pm}-r_2)}}\times\\
   &\frac{\log\left( \frac{\big(D_{\pm} \sqrt{1-D_{\pm}^2k_r^2} + \sqrt{1-k_r^2\snr^2}\snr\big)^2+ \big(k_r(D_{\pm}^2-\snr^2)\big)^2}{\big(D_{\pm} \sqrt{1-D_{\pm}^2k_r^2} -\sqrt{1-k_r^2\snr^2}\snr\big)^2+ \big(k_r(D_{\pm}^2-\snr^2)\big)^2}\right)
    }{\sqrt{(A^2(r_{\pm}-r_2)-(r_1-r_{\pm})(r_2^2-B^2+r_1r_{\pm}-r_2(r_1+r_{\pm}))}}. \;\;\;\;\;
    \end{aligned}
\end{equation}
Where we have defined,
\begin{align}
    D_{\pm} &= \frac{\sqrt{4AB(r_1-r_{\pm})(r_{\pm}- r_2)}}{A(r_{\pm}-r_2)+B(r_1-r_{\pm})}\text{ , and}\\
	\xi_r(\lam) &= \am(\sqrt{ (1-\EN^2) A B} \lam|k_r^2).\label{eq:xiR}
\end{align}
In the above we have suppressed the explicit dependence of $\xi_r$ on $\lam$ in the interest of readability. The $\arctan$ function seen in Eq.~\eqref{eq:Ir2} is the two argument arctan function which tracks the sectors of the numerator and denominator.
\subsection{Solutions to Equations of Motion}
The solution for Eqs.~\eqref{eq:MINOsolution}-\eqref{eq:Irpm} for the elliptic integrals can then be readily substituted back into Eq.~\eqref{eq:reducedformphi } and  Eq.~\eqref{eq:reducedformt} giving the full form of the solutions to generic plunges. We present all solutions parameterised in terms on Mino time $(\lam)$, which is done by inverting the solution to Eq.~\eqref{eq:MINOsolution} to obtain $r(\lam)$ then substituting the solution for $r(\lam)$ everywhere $r$ appears in  Eqs.~\eqref{eq:Ir}-\eqref{eq:Irpm} . These solutions are provided in a fully analytic form. 
The radial equation is first found by inverting the solution for Eq.~\eqref{eq:MINOsolution} to give,
\begin{equation}
r(\lam) = \frac{(A-B)(Ar_2-Br_1)\snr^2+2A B(r_2+r_1)-2AB(r_1-r_2)\cnr}{4 A B + (A-B)^2 \snr^2}.
\end{equation}
The solutions to the polar equation remain the same as for the ISSO case. 
    \begin{figure}[t!]
    \centering
    \includegraphics[width=100mm]{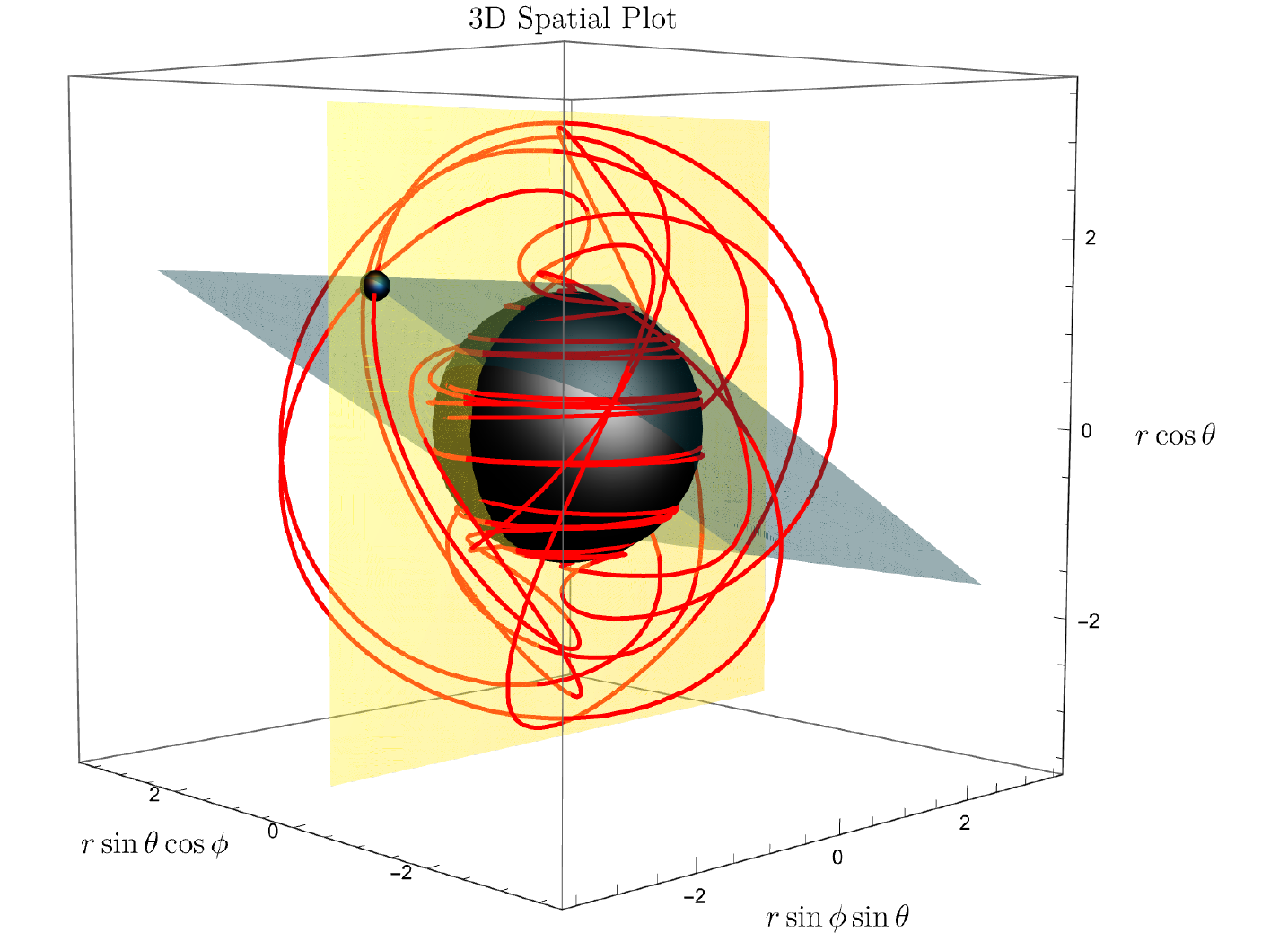}
         \caption{Orbital plot of a generic plunging geodesic with parameter values $(a,\EN,\ANG,Q) = (0.9,0.94,0.1,12)$ and $\theta = \mathrm{arccos}(z)$. The larger black sphere gives the horizon of the black hole where as the smaller sphere simply gives a point along the geodesics. Here the cyan and orange planes are the azimuthally co-precessing and poloidally co-rotating planes respectively. }
    \label{fig:genericplots}
\end{figure}
Taking Eq.~\eqref{eq:reducedformphi }, the solution  to the azimuthal equations of motion can then immediately be found from the solutions of Eq.~\eqref{eq:ellipticBasis}
\begin{align}
    \phi(\lam) = \phi_r(\lam) + \phi_z(\lam) - a \EN \lam.
\end{align}   
Similarly, from Eq.~\eqref{eq:reducedformt} we can now obtain
\begin{align}
    t(\lam) = t_r(\lam) +   t_z(\lam) +a \ANG \lam,
\end{align}
where both $\phi_z$ and $t_z$ can be taken from the ISSO case. The solution for proper time as a function of Mino time can also be found as, 
\begin{align}
     \tau(\lam) =& \tau_r(\lam)+ \tau_z(\lam), \text{ with}
\\
    \tau_r(\lam) =&\mathcal{I}_{r^2}(\lam) , \text{ and}
\\
    \tau_z(\lam) =&\frac{ z_2 }{(1-\EN^2)}\left( F(\amz|k_z^2 )-E(\amz|k_z^2)\right).
\end{align}
Having obtained the full set of solutions for generic plunges we plot the spatial component depicting the orbital evolution of the generic plunging geodesics Fig.~\ref{fig:genericplots}. 
\begin{figure}[t!]
    \centering
    \includegraphics[width=150mm]{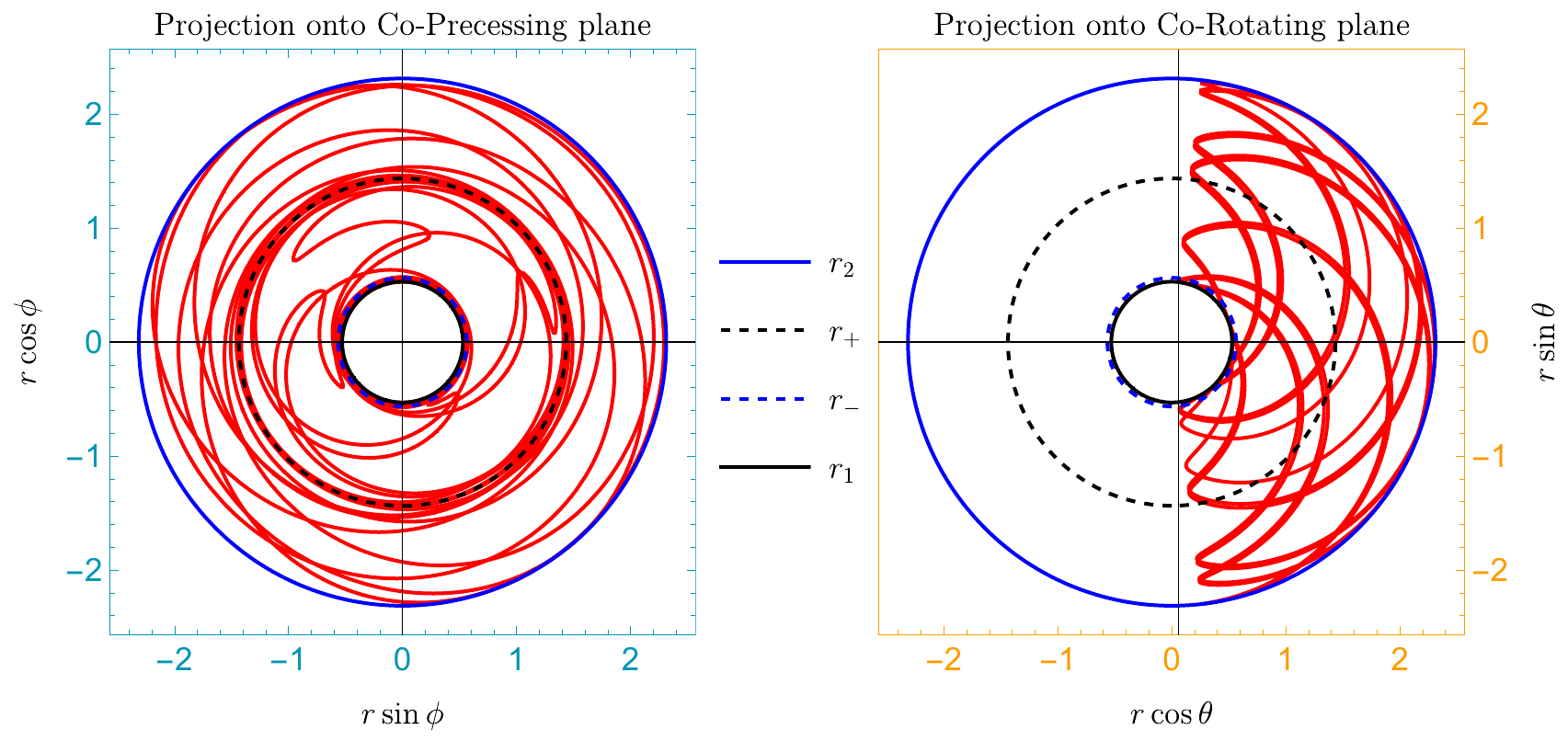}
    \caption{Co-rotating orbital plots of a generic plunging geodesic with parameter values $(a,\EN,\ANG,Q) = (0.9,0.94,0.1,12)$. Left Image: orthogonal projection of generic plunge onto the co-precessing azimuthal plane. Right Image: orthogonal projection of generic plunge onto co-rotating polar plane. The cyan and orange frames represent the projection onto the planes seen in \ref{fig:genericplots} as they track a particle following the geodesic trajectory.}
    \label{fig:coorotatinggeneric}
\end{figure}
More informatively, the orthogonal projection onto the azimuthally co-precessing and poloidally co-rotating planes in Fig.~\ref{fig:coorotatinggeneric} again show the divergences at either horizon in the $\phi$ coordinate. Importantly, once we have constructed these solution, we check that $\EN$ and $\ANG$ are indeed still conserved by evaluating $ -u^{\nu}g_{\mu\nu}\left(\frac{\partial}{\partial t}\right)^{\nu}$ and $ u^{\nu}g_{\mu\nu}\left(\frac{\partial}{\partial \phi}\right)^{\nu} $ explicitly. We confirm this is the case for all values of $\lam$, not only acting as a consistency check of our equations, but also showing we have selected the correct branches of the solution between each of the horizon divergences. As a final consistency check, we substitute our solutions back into the equations of motion for both the ISSO and generic case and find that our solutions do in fact solve the original equations. Finally, from our solutions the radial and polar frequencies with respect to Mino time for a generic plunge can be obtained, 
\begin{align}
    \Upsilon_r &=   \frac{ \pi \sqrt{ A B(1-\EN^2)}}{2 \elK(k_r^2)}, \text{ and }\\
    \Upsilon_z &= \frac{\pi z_2 }{2  \elK(k_z^2)  }.
\end{align}

\section{Discussion}\label{sec:conclusion}
In this work we have obtained the closed form analytic solutions for generic bound plunging geodesics in Kerr. We have also paid particular attention to the special edge case of plunges starting asymptotically from the innermost stable precessing circular orbit, in which the solutions take a particularly simple form. This generalises the result of~\cite{Mummery:2022ana} to inclined motion. The general expression for the inflow in this case, is more involved due to oscillations coming from the polar motion. Therefore we have provided two simplified approximations for the inflow rate. We have also found that the geodesics asymptoting from the ISSO can be parameterised purely in terms of black hole spin and the radius of the ISSO. We expect these solutions to have applications in the modelling of accretion flow in Kerr spacetimes. In addition to the special case of ISSO plunges, we expect the provided generic solutions for plunging geodesics to find practical use in modelling the inspiral of binary black holes, since in the small mass-ratio limit they will describe the final phase of the inspiral before merger. As small mass-ratio methods are applied to more equal mass systems, including this phase becomes increasingly important. For ease of application we have incorporated our results in the {\tt KerrGeodesics} package in the Black hole perturbation toolkit~\cite{BHPToolkit}.

In this work we have restricted attention to ``bound'' plunging geodesics, i.e. geodesics with $\EN<1$. In principle, the explicit solutions given here can easily be extended to the case of ``direct plunges'' coming from infinity and falling into the black hole, or deeply bound plunges inside a scattering geodesic. In both cases, this involves taking the expressions in this paper in terms of the outermost real root and analytically continuing it past positive infinity to negative values (e.g. by considering the reciprocal root). However, this will not give all  $\EN>1$ plunging geodesics, which also includes so-called ``vortical'' geodesics with negative values of $Q$~\cite{Compere:2020eat} leading to qualitatively different behaviour of the polar solutions.

\section*{Acknowledgements}
The authors thank Daniel D'Orazio for useful conversations in relation to potential astrophysical applications of these results and David Pereñiguez regarding helpful comments on the final draft of this paper.
We acknowledge support from the Villum Investigator program supported by the VILLUM Foundation (grant no. VIL37766) and the DNRF Chair program (grant no. DNRF162) by the Danish National Research Foundation.
This work makes use of the Black Hole Perturbation Toolkit~\cite{BHPToolkit}.

\section*{References}
\bibliography{references}

\appendix
\section{Solutions for timelike geodesics asymptoting from an unstable spherical orbit}\label{A:USO}
In this appendix, we present the solutions for plunging geodesics asymptoting from a generic USO. In the phase space of geodesics, the USOs denote the separatrix between eccentric inclined bound orbits and generic plunges. As such, they play a key role when considering the transition of generic inspirals to plunge. As in sections \ref{sec:ISSO} and  \ref{sec:deeplybound} the components of the solutions depending on the polar angle are left unchanged by restricting to this special case so we are only required to solve for $r(\lam),\; \phi_r(\lam)$ and $t_r(\lam)$ as described in the notation of the main body of the text. In particular, plunges from an USO occur when the radial potential obtains a double root as opposed to a triple root (which arises in the ISSO case). Explicitly the USO plunge occurs when the radial potential admits a form,
\begin{equation}
	R(r) = (1-\EN^2)(r-r_2)(r_1-r)(r_s-r)^2\, , 
\end{equation}
with $r_2 < r_{-} < r_+ < r_s < r_1$. Where $r_s$ is the radius of the USO. Specifically, the solutions we present provide bound motion for $r \in (r_2 , r_s)$. The Penrose diagram depicting this class of motion is the same as in Fig.~\ref{fig:penrose} for the ISSO plunges. As we have imbued the radial potential with a double root we have reduced one degree of freedom in our systems parameterisation. A natural choice of parameterisation for USO plunges is then given by $\{a,r_s, Q\}$. Teo~\cite{Teo:2020sey} has determined expressions for $\EN$ and $\ANG$ as functions of $\{a,r_s, Q\}$ for the case of USOs which are given as,
\begin{align}
	\EN &= \frac{r_s^3(r_s - 2) - a (a Q -\sqrt{A})}{r_s^2 \sqrt{r_s^3 (r_s - 3) - 2 a (a Q - \sqrt{A})}}\, ,\\
	\ANG &= - \frac{2 M r_s^3 + (r_s^2 + a^2)(a Q -\sqrt{A})}{r_s^2 \sqrt{r_s^3 (r_s - 3) - 2 a (a Q - \sqrt{A})}}\, , \intertext{where}\\ \notag
	A &= r_s^5 - Q (r_s - 3)r_s^3 + a^2 Q^2.
\end{align}
As USO plunges asymptote from the USO they must share the same constants of motion. Similarly to the case of the ISSO the reduced complexity in the root structure of the radial potential allows one to find solutions to the radial integrals fully in terms of elementary functions of Mino time.  Following the procedure shown explicitly in section \ref{sec:ISSO}  and defining, 
\begin{equation}
 \xi_r(\lam) = \frac{1}{2} \lambda  \sqrt{1-\mathcal{E}^2} \sqrt{r_1-r_s} \sqrt{r_s-r_2}\,,
 \end{equation}
the solution for the radial coordinate is found to be given by,
\begin{equation}
r(\lam) =\frac{r_2 \left(r_1-r_s\right)+r_1 \left(r_s-r_2\right) \tanh ^2\left( \xi_r\right)}{r_1-r_s+\left(r_s-r_2\right) \tanh ^2\left( \xi_r\right)}.
\end{equation}
Writing the azimuthal equation in the form $\phi(\lam) = \phi_r(\lam)+ \phi_z(\lam)  - a \EN \lam$  and the time equation in the form $t(\lam) = t_r(\lam)+ t_z(\lam)  + a \ANG \lam$  we find that for the case of timelike geodesics asymptoting from an USO,
\begin{align}
&\text{$\phi_r $}(\lambda )=\frac{2 a }{\sqrt{1-\mathcal{E}^2}}\Bigg(\frac{\lambda  \sqrt{1-\mathcal{E}^2} \left(a^2 \mathcal{E}-a \mathcal{L}+\mathcal{E} r_s^2\right)}{2 \left(r_s-r_-\right) \left(r_s-r_+\right)}\\ \notag
&+\frac{\left(a^2 \mathcal{E}-a \mathcal{L}+r_-^2 \mathcal{E}\right) \log \left(\frac{\left(\sqrt{\left(r_- -r_2\right) \left(r_1-r_s\right)}+\sqrt{\left(r_1-r_-\right) \left(r_s -r_2\right)} \tanh \left( \xi_r\right)\right){}^2}{\left(\sqrt{\left(r_- - r_2\right) \left(r_1-r_s\right)}-\sqrt{\left(r_1-r_-\right) \left(r_s - r_2\right)} \tanh \left( \xi_r\right)\right){}^2}\right)}{4 \left(r_--r_+\right) \sqrt{r_1-r_-} \sqrt{r_- -r_2} \left(r_--r_s\right)}\\ \notag
&+\frac{\left(a^2 \mathcal{E}-a \mathcal{L}+r_+^2 \mathcal{E}\right) \log \left(\frac{\left(\sqrt{\left(r_- -r_2\right) \left(r_1-r_s\right)}+\sqrt{\left(r_1-r_+\right) \left(r_s-r_2\right)} \tanh \left( \xi_r\right)\right)^2}{\left(\sqrt{\left(r_+ -r_2\right) \left(r_1-r_s\right)}-\sqrt{\left(r_1-r_+\right) \left(r_s-r_2\right)} \tanh \left( \xi_r\right)\right){}^2}\right)}{4 \left(r_+-r_-\right) \sqrt{r_1-r_+} \sqrt{r_+-r_2} \left(r_+-r_s\right)}\Bigg),\\ \notag
\intertext{and, }
&t_r(\lambda )=\frac{1}{\sqrt{1-\mathcal{E}^2}}\Bigg(\frac{\lambda  \sqrt{1-\mathcal{E}^2} \left(a^2+r_s^2\right) \left(a^2 \mathcal{E}-a \mathcal{L}+\mathcal{E} r_s^2\right)}{\left(r_s-r_-\right) \left(r_s-r_+\right)}\\ \notag
&+\frac{\left(a^2+r_-^2\right) \left(a^2 \mathcal{E}-a \mathcal{L}+r_-^2 \mathcal{E}\right) \log \left(\frac{\left(\sqrt{r_--r_2} \sqrt{r_1-r_s}+\sqrt{r_1-r_-} \sqrt{r_s-r_2} \tanh \left( \xi_r\right)\right){}^2}{\left(\sqrt{r_--r_2} \sqrt{r_1-r_s}-\sqrt{r_1-r_-} \sqrt{r_s-r_2} \tanh \left( \xi_r\right)\right){}^2}\right)}{2 \left(r_--r_+\right) \sqrt{r_1-r_-} \sqrt{r_--r_2} \left(r_--r_s\right)}\\ \notag
&+\frac{\left(a^2+r_+^2\right) \left(a^2 \mathcal{E}-a \mathcal{L}+r_+^2 \mathcal{E}\right) \log \left(\frac{\left(\sqrt{r_+-r_2} \sqrt{r_1-r_s}+\sqrt{r_1-r_+} \sqrt{r_s-r_2} \tanh \left( \xi_r\right)\right){}^2}{\left(\sqrt{r_+-r_2} \sqrt{r_1-r_s}-\sqrt{r_1-r_+} \sqrt{r_s-r_2} \tanh \left( \xi_r\right)\right){}^2}\right)}{2 \left(r_+-r_-\right) \sqrt{r_1-r_+} \sqrt{r_+-r_2} \left(r_+-r_s\right)}\\ \notag
&+\frac{\left(r_1-r_2\right) \mathcal{E} \sqrt{\left(r_1-r_s\right) \left(r_s-r_2\right)} \tanh \left( \xi_r\right)}{-\left(\left(r_2-r_s\right) \tanh ^2\left( \xi_r\right)\right)-r_s+r_1}\\ \notag
&-\mathcal{E} \left(2 \left(r_s+r_-+r_+\right)+r_1+r_2\right) \tan ^{-1}\left(\frac{\sqrt{r_s-r_2} \tanh \left( \xi_r\right)}{\sqrt{r_1-r_s}}\right)\Bigg). \notag
\end{align}
Where $\phi_z \text{ and } t_{z}$ are left unchanged from Eqs.~\ref{eq:polarphi} and \ref{eq:polartime}. In the equatorial limit ($Q=0$) these solutions agree with the equatorial plunging orbits described in section V.C of~\cite{Mummery:2023hlo}, where the radius of the unstable circular orbit is chosen between the innermost bound circular orbit (IBCO) and the ISCO.

\end{document}